\documentclass[lettersize,journal]{IEEEtran}

\usepackage{hyperref}
\usepackage{url}

\usepackage{amsmath,amsfonts}
\usepackage{algorithmic}
\usepackage{algorithm}
\usepackage{array}
\usepackage[caption=false,font=normalsize,labelfont=sf,textfont=sf]{subfig} 
\usepackage{textcomp}
\usepackage{stfloats}
\usepackage{url}
\usepackage{verbatim}
\usepackage{graphicx}
\usepackage{booktabs}
\usepackage{comment}
\usepackage{multirow}
\usepackage{colortbl}
\usepackage{xcolor}
\usepackage{bbm}

\definecolor{lightgreen}{RGB}{200, 255, 200}
\definecolor{medgreen}{RGB}{140, 230, 140}
\definecolor{darkgreen}{RGB}{90, 200, 90}
\definecolor{tableblue}{RGB}{225, 235, 252}

\usepackage{arydshln}

\usepackage{orcidlink}

\title{SelfTTS: Cross-Speaker Style Transfer through Explicit Embedding Disentanglement and Self-Refinement using Self-Augmentation}

\author{Lucas~H.~Ueda~\orcidlink{0000-0002-1029-3420},
        João~G.~T.~Lima~\orcidlink{0009-0006-6336-8808},
        Pedro R. Corrêa~\orcidlink{0000-0002-3411-0959},
        Flávio O. Simões~\orcidlink{0000-0001-5104-1222}, 
        Mário U. Neto~\orcidlink{0009-0008-7254-285X}
        and~Paula~D.~P.~Costa~\orcidlink{0000-0002-1534-5744}%
\thanks{Lucas H. Ueda, João G. T. Lima, Pedro R. Corrêa  and Paula D. P. Costa are with the Department of Computer Engineering and Automation (DCA), Faculdade de Engenharia Elétrica e de Computação, and the AI Lab., Recod.ai, Institute of Computing, Universidade Estadual de Campinas (UNICAMP), Campinas, Brazil (e-mail: l156368@dac.unicamp.br, j237473@dac.unicamp.br,  p243236@dac.unicamp.br, paulad@unicamp.br).}%
\thanks{Flávio O. Simões and Mário U. Neto are with the CPQD, Campinas, Brazil (e-mail: simoes@cpqd.com.br, uliani@cpqd.com.br).}%
\thanks{Paula D. P. Costa (paulad@unicamp.br) is the corresponding author.}%
\thanks{This work is partially funded by CAPES – Finance Code 001. It is also supported by FAPESP (BI0S \#2020/09838-0 and Horus \#2023/12865-8). This project was supported by MCTI under Law 8.248/1991, PPI-Softex, published as Cognitive Architecture (Phase 3), DOU 01245.003479/2024-1.}
}

\begin{document}

\maketitle

\begin{abstract}

    This work presents SelfTTS, a text-to-speech (TTS) model designed for cross-speaker style transfer that eliminates the need for external pre-trained speaker or emotion encoders. The architecture achieves emotional expressivity in neutral speakers through an explicit disentanglement strategy utilizing Gradient Reversal Layers (GRL) combined with cosine similarity loss to decouple speaker and emotion information. We introduce Multi Positive Contrastive Learning (MPCL) to induce clustered representations of speaker and emotion embeddings based on their respective labels. Furthermore, SelfTTS employs a self-refinement strategy via Self-Augmentation, exploiting the model's voice conversion capabilities to enhance the naturalness of synthesized speech. Experimental results demonstrate that SelfTTS achieves superior emotional naturalness (eMOS) and robust stability in target timbre and emotion compared to state-of-the-art baselines.
    
\end{abstract}

\section{Introduction}

Generating expressive speech in text-to-speech (TTS) models typically requires recordings that capture specific stylistic nuances. However, obtaining such high-quality data for every speaker is often unfeasible. Recent literature addresses this limitation through cross-speaker style transfer, where the prosodic characteristics of a reference audio signal are mapped onto a target speaker who has provided only neutral recordings.

A common framework involves a style encoder module, which extracts a style representation embedding from reference audio to condition the TTS prosody. This encoder generally derives an embedding representation from a reference mel-spectrogram with the goal of creating a style space that isolates prosody from linguistic content and speaker identity. In scenarios involving labeled datasets this style space can be further used for post-hoc analysis and for conditioning the TTS inference for cross-speaker style transfer~\cite{kwon2019emotional, sorin2020principal}.

Several studies incorporate an emotion classification layer at the output of the style encoder to improve the emotional alignment of these representations~\cite{im2022emoqtts, zhou2023speech}. Alternatively, Kulkarni et al. (2021)~\cite{kulkarni2021improving} employ an N-Pair loss to generate a style space that is more effectively clustered across different emotions, while Ngo et al. (2024)~\cite{ngo2024learning} utilizes Supervised Contrastive Loss. More recent works, such as Gudmalwar et al. (2024)~\cite{gudmalwar2024vecl}, utilize separate pre-trained models to extract distinct embeddings for both speaker identity and style.

In practice, however, these modules often suffer from speaker leakage, a condition where the reference speaker's timbre is erroneously captured~\cite{skerryryan2018towards, valles-perez2021improving, ueda2021are}. This leads to degraded performance and identity mismatch during cross-speaker inference~\cite{an2022disentangling}.

To achieve better disentanglement between emotion and speaker embeddings, some studies incorporate adversarial training with Gradient Reversal Layer~\cite{ganin2015unsupervised} (GRL). In this setup, a classifier identifies undesired labels while the GRL reverses the gradients during the backward pass, preventing the model from encoding information related to those labels. Although this technique is widely used in modern TTS models~\cite{shang2021incorporating, zaidi2022daft}, relying solely on external labels can be noisy and may not explicitly handle the entanglement inherent in the embeddings conditioning the model. Despite these efforts, maintaining high naturalness in cross-speaker scenarios remains a significant challenge~\cite{sigurgeirsson2023prosody}.

Alternatively, timbre perturbation techniques, such as formant shifting, have been employed to force the model to decouple emotion and prosody from the speaker's vocal characteristics~\cite{choi2021neural, lei2022cross, zhu2024metts}. Other approaches rely on filtering only the first 20 mel-bins of the input mel-spectrogram for the style encoder, as this range contains significant prosodic information with minimal timbre or content interference compared to the full band~\cite{ren2022prosospeech, jiang2023megatts}.

When expressive data is limited, synthetic data generation offers a viable path forward. This often involves using a Voice Conversion (VC) model to transform an expressive source speaker into the target speaker’s voice, followed by fine-tuning a TTS model on this augmented data~\cite{huybrechts2021lowresource, ribeiro2022cross}. Recently, the use of synthetic samples to improve disentanglement and naturalness has been further explored~\cite{ueda2024xploring}. Leveraging the VC capabilities of VITS~\cite{kim2021conditional}, the E3-VITS model~\cite{jung2023e3vits} proposed batch-permuted style perturbation to improve style transfer using unpaired style samples generated by the model itself.

Synthetic data has also been extended to multilingual systems to retain content across languages~\cite{yoon2024enhancingmultilingual}. Nevertheless, the overall quality of the resulting TTS is strictly bounded by the quality of the synthetic samples, especially when handling high-intensity expressive data~\cite{terashima2022cross}.

In this work, we present SelfTTS, a TTS model capable of transferring emotion to neutral speakers without requiring external pre-trained speaker or emotion encoders. The proposed model adopts an explicit disentanglement strategy using a GRL combined with a cosine similarity loss to remove speaker and emotion information from specific layers. We also propose the use of Multi Positive Contrastive Learning (MPCL) loss~\cite{tian2023stablerep} to induce both speaker and emotion embeddings to produce clustered representations based on their respective labels. Furthermore, we adopt a self-refinement strategy with Self-Augmentation, utilizing the inherent voice conversion capabilities of the VITS-based architecture to enhance the naturalness of generated speech while maintaining emotional expressivity and target timbre. Audio samples are available \url{https://ai-unicamp.github.io/publications/tts/selftts/}. The contributions of this work are validated through a detailed experimental setting and include:

\begin{enumerate}
    \item A model utilizing MPCL loss to generate effective label-oriented clusters for both speaker and emotion embeddings without the need for complex batching pipelines;
    \item An explicit disentanglement framework using cosine similarity loss in conjunction with GRL to effectively remove overlapping information from embeddings, thereby enabling robust cross-speaker style transfer;
    \item The use of the model's own Voice Conversion capability for Self-Augmentation to improve the naturalness of synthesized speech;
    \item A comprehensive evaluation, including cross-corpus experiments, conducted using open datasets and publicly available code\footnote{\url{https://github.com/AI-Unicamp/SelfTTS}}.
\end{enumerate}

\section{Methodology}

In this section, we will present the methodology adopted in this work.

\subsection{SelfTTS}

\begin{figure*}[ht]
    \centering

    \includegraphics[width=1.0\textwidth]{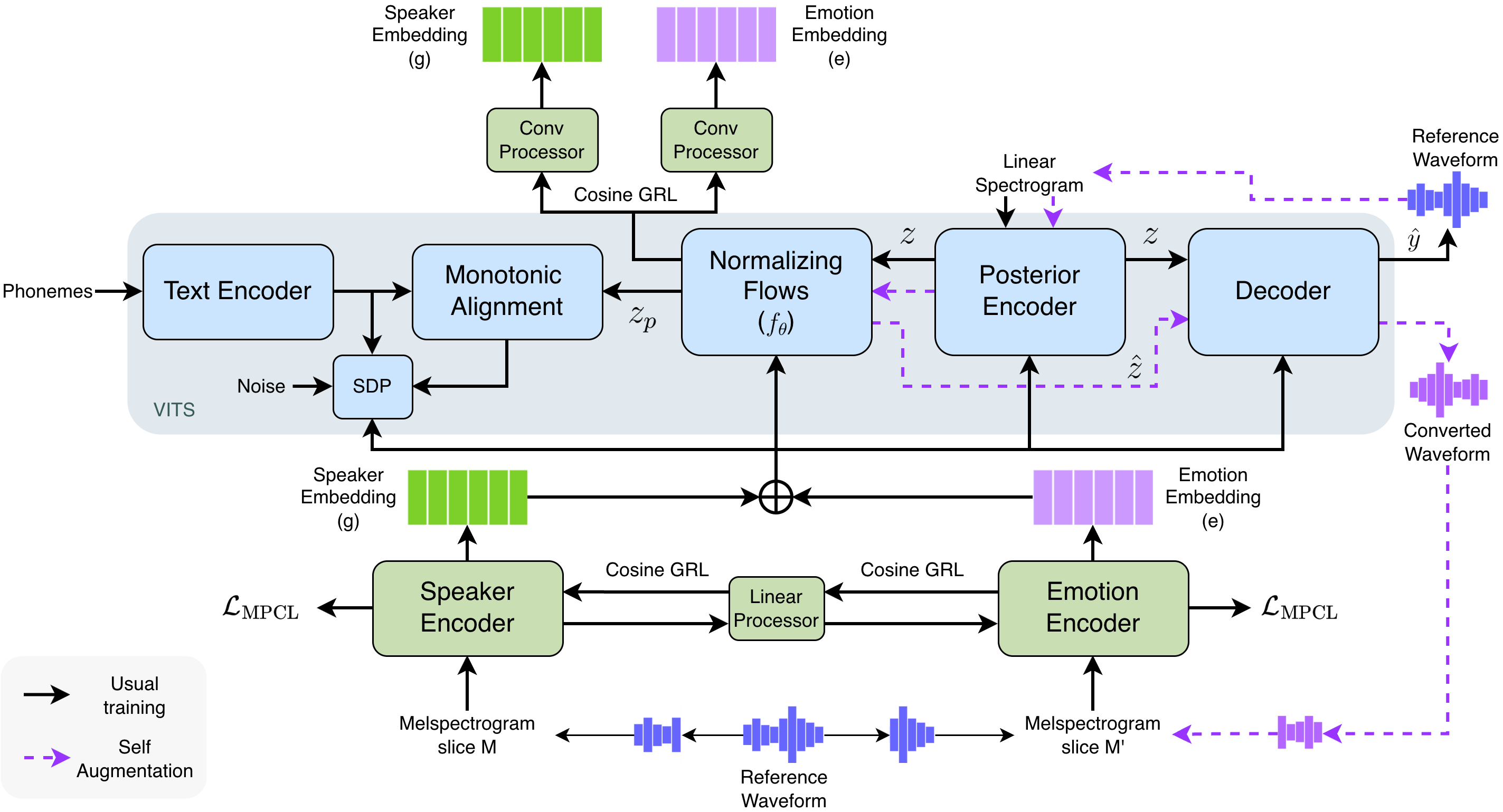}
    \caption{SelfTTS model architecture. The emotion and speaker encoders receive mel-spectrogram slices of the reference waveform as input. Each encoder is optimized using the MPCL loss, while their embeddings are disentangled through a cosine-based GRL applied on top of the Linear Processor output for each encoder. The final forward step of the Normalizing Flows ($z_p$) is also disentangled using a cosine-based GRL, through a Convolutional Processor that predicts the corresponding emotion or speaker embedding. SDP stands for \textit{Stochastic Duration Predictor}. Purple dashed arrows indicate the proposed Self-Augmentation pipeline.}
    \label{fig:selftts_architecture}

\end{figure*}

The proposed SelfTTS is built upon the Variational Inference with adversarial learning for end-to-end Text-to-Speech (VITS)~\cite{kim2021conditional} framework. We extend the original conditional Variational Autoencoder (CVAE) structure by integrating dedicated speaker ($g$) and emotion ($e$) embeddings across all major components. Specifically, the \textit{Posterior Encoder}, \textit{Residual Coupling Blocks} (Normalizing Flows), \textit{Stochastic Duration Predictor} (SDP), and the \textit{Waveform Decoder} are all conditioned on these embeddings. The SelfTTS architecture is represented in Figure~\ref{fig:selftts_architecture}.

The \textit{Posterior Encoder} processes the linear spectrogram $x_{lin}$ of the target speech to parameterize an approximate posterior distribution $q_{\phi}(z|x_{lin}, g, e)$. A latent variable $z$ is sampled and transformed into the prior space via a series of four invertible \textit{Residual Coupling Blocks} that constitute the Normalizing Flows $f_{\theta}$. This transformation yields the style-neutral and speaker-agnostic representation $z_p = f_{\theta}(z; g, e)$, which is aligned with the output of the \textit{Text Encoder} by minimizing the Kullback-Leibler (KL) divergence:

\begin{equation}
    L_{kl} = \log q_{\phi}(z|x_{lin}, g, e) - \log p_{\theta}(z_p | c_{text}, A)
\end{equation}

where $A$ represents the hard monotonic attention matrix estimated via Monotonic Alignment Search (MAS).

To synthesize the raw waveform, the \textit{Waveform Decoder} upsamples the latent representation $z$. To maintain computational efficiency, we employ windowed generator training, where only a random segment of the latent sequence $z_{slice}$ is decoded into a partial waveform $\hat{y}$. The speaker and emotion embeddings are generated using a Reference Encoder~\cite{skerryryan2018towards} (RE) architecture. The RE consists of a 6-layer convolutional neural network that processes a mel-spectrogram reference signal, followed by a Gated Recurrent Unit (GRU) to summarize the sequence into a single fixed-length embedding. As input each encoder receives a random slice of the reference audio to also make it robust against content leakage~\cite{chen2022finegrained}. Each slice is randomly selected in a window of random size between the whole sample length and half of it to align the training design to inference time where usually the sentence is longer than the emotional reference. An explicit embedding disentanglement technique is applied to these vectors to decouple speaker and emotion information, ensuring the model can transfer emotions even to neutral-only speakers. 

\subsubsection{Multi Positive Contrastive Learning}

To enforce each encoder to generate clustered representations for both speaker and emotion labels we adopt the Multi Positive Contrastive Learning (MPCL) loss~\cite{tian2023stablerep}.

We define a contrastive categorical distribution \( q \) that measures the relative compatibility between an anchor \( a \) and each candidate \( b_i \):

\begin{equation}
    q_i = \frac{\exp\left( \frac{\mathbf{a} \cdot \mathbf{b}_i}{\tau} \right)}
    {\sum_{j=1}^{K} \exp\left( \frac{\mathbf{a} \cdot \mathbf{b}_j}{\tau} \right)}
    \label{eq:softmax_mpcl}
\end{equation}

where \( \tau \in \mathbb{R}_+ \) is a temperature parameter that controls the sharpness of the distribution, and both \( a \) and \( b_i \) are assumed to be \(\ell_2\)-normalized.

The target categorical distribution \( c \) is defined as:

\begin{equation}
c_i =
\frac{\mathbbm{1}_{\text{match}}(a, b_i)}
{\sum_{j=1}^{K} \mathbbm{1}_{\text{match}}(a, b_j)},
\end{equation}

where \( \mathbbm{1}_{\text{match}}(\cdot,\cdot) \) equals one if the anchor and candidate share the same emotional label, and zero otherwise.

The MPCL loss is then defined as the cross-entropy between the target distribution \( c \) and the predicted distribution \( q \):

\begin{equation}
\mathcal{L}_{\mathrm{MPCL}} = - \sum_{i=1}^{K} c_i \log q_i.
\end{equation}

The MPCL loss doesn't require specific batch design and can be trained in an end-to-end fashion within the SelfTTS model.

\subsubsection{Explicit Embedding Disentanglement}

Although widely used, standard techniques like vanilla CE-based Gradient Reversal Layer~\cite{ganin2015unsupervised} (GRL) do not necessarily prevent speaker leakage within the emotion encoder. Consider a case where only one speaker in the dataset provides expressive speech, by design, a correlation exists between emotion labels and speaker labels, therefore vanilla CE-based GRL is not capable of removing speaker information using external speaker labels. To address this, we propose applying a GRL directly between the embeddings generated by the emotion and speaker encoders. We employ cosine similarity as the loss function between a simple linear processor of these embeddings ($\Phi_{linear}(e)$ and $\Phi_{linear}(g)$) and the corresponding other embedding. Consequently, the model does not rely on labels during the GRL process, it propagates the reversed gradient based on the embeddings themselves, explicitly forcing disentanglement between speaker and emotion representations. The $\Phi_{linear}$ consists of a stack of three linear layers with ReLU activation outputting the same dimension as the emotion and speaker embeddings. During training, the proposed model learns to generate clustered emotion and speaker representations of their respective labels via MPCL loss while simultaneously ensuring these representations remain disentangled.

The cosine similarity loss $\mathcal{L}_{\text{cos\_emb}}$ used to disentangle both the representations is defined as the two equations below:

\begin{equation}
\mathcal{L}_{\text{cos\_emb}}(g, e) = \frac{\Phi_{linear}(g) \cdot e{.\text{detach}()} }{\lVert \Phi_{linear}(g) \rVert \cdot \lVert e{.\text{detach}()}  \rVert}
\end{equation}

\begin{equation}
\mathcal{L}_{\text{cos\_emb}}(e, g) = \frac{\Phi_{linear}(e) \cdot g{.\text{detach}()} }{\lVert \Phi_{linear}(e) \rVert \cdot \lVert g{.\text{detach}()} \rVert}
\end{equation}

We also enforce robust disentanglement at the latent level $z_p$. While the \textit{Text Encoder} contains purely linguistic information, the \textit{Posterior Encoder} is susceptible to leaking speaker and emotional characteristics because it receives $x_{lin}$ as input. Similarly to speaker and emotion embeddings, we apply the GRL to $z_p$ coupled with two convolutional processors, a speaker processor ($\Phi_{conv}(g)$) and an emotion processor ($\Phi_{conv}(e)$), to ensure that the representation $z_p$ remains invariant to speaker and emotion. These convolutional processors consist of three 1D convolutional layers with reLU activation function and a mean pooling head that generates the same dimension as emotion and speaker encoder outputs. The GRL penalizes the presence of such identifiable information by reversing gradients during backpropagation, forcing the flows to produce a truly style-neutral and speaker-agnostic latent space $z_p$ where speaker and emotion characteristics are only reintroduced during the reverse flow or in subsequent modules. Equations~\ref{eq:cos_content_e} and \ref{eq:cos_content_g} represent the content disentanglement losses.

\begin{equation}
\mathcal{L}_{cos\_{content}}(z_{\text{p}}, e) = \frac{\Phi_{conv}(z_{\text{p}}) \cdot e{.\text{detach}()} }{\lVert \Phi_{conv}(z_{\text{p}}) \rVert \cdot \lVert e{.\text{detach}()}  \rVert}
\label{eq:cos_content_e}
\end{equation}

\begin{equation}
\mathcal{L}_{cos\_{content}}(z_{\text{p}}, g) = \frac{\Phi_{conv}(z_{\text{p}}) \cdot g{.\text{detach}()} }{\lVert \Phi_{conv}(z_{\text{p}}) \rVert \cdot \lVert g{.\text{detach}()}  \rVert}
\label{eq:cos_content_g}
\end{equation}

\subsubsection{Self-Augmentation}

The voice conversion capability is derived from the invertible nature of Normalizing Flows and the disentanglement of the latent space. To perform conversion, a source audio is mapped to a latent variable $z$ via the posterior encoder $q_{\phi}(z|x_{lin}, g_{src}, e_{src})$. The forward flow $f_{\theta}$ then transforms $z$ into a style-neutral and speaker-agnostic representation $z_p = f_{\theta}(z; g_{src}, e_{src})$, which resides in a prior distribution containing only linguistic content. Conversion is achieved by passing $z_p$ through the inverse flow $f_{\theta}^{-1}$ conditioned on target identity and emotion: 

\begin{equation}
    \hat{z} = f_{\theta}^{-1}(z_p; g_{tgt}, e_{tgt})
\end{equation}

The resulting latent variable $\hat{z}$ is then processed by the decoder $G$ to synthesize the final waveform $\hat{y} = G(\hat{z}; g_{tgt}, e_{tgt})$, effectively transferring the target's voice and emotion while preserving the source's utterance.

Building upon this, we propose leveraging the model’s voice conversion capability optimized by our disentanglement techniques to generate synthetic data for self-refinement via Self-Augmentation. 

The Self-Augmentation process involves randomly permuting the speaker references to generate synthetic samples that maintain the original emotion across different voices ($\hat{y} = G(\hat{z}; g_{tgt}, e_{src})$). Although a similar permutation could be applied to emotional references while keeping the speaker identity fixed, we observe that since the model's conversion process does not explicitly control duration, emotions such as ``Sad'' could be negatively impacted by the lack of adjustment in speaking rate. A fixed proportion of the batch is masked and replaced with these generated synthetic samples. The mixed batch with both ground-truth and generated samples becomes the input of the emotion encoder, providing different speaker voices for the same emotions. This Self-Augmentation stage is applied only after the base architecture has been pre-trained, ensuring the model has already obtained sufficient cross-speaker style transfer capability before focusing on further performance refinement.

The model is optimized using a combination of VITS original losses augmented with all the proposed new losses (Equation~\ref{eq:total_loss}).

\begin{equation}
\begin{split}
    \mathcal{L}_{total} = \overbrace{\mathcal{L}_{recon} + \mathcal{L}_{kl} + \mathcal{L}_{dur} + \mathcal{L}_{adv}(G) + \mathcal{L}_{fm}(G)}^{\text{original VITS losses}} \\
    + \mathcal{L}_{\text{cos\_content}}(z_{\text{p}}, e) +    \mathcal{L}_{\text{cos\_content}}(z_{\text{p}}, g) \\
    + \mathcal{L}_{\text{cos\_emb}}(g, e) +
    \mathcal{L}_{\text{cos\_emb}}(e, g) +\mathcal{L}_{\text{MPCL}}(e) + \mathcal{L}_{\text{MPCL}}(g)
\end{split}
\label{eq:total_loss}
\end{equation}

\begin{figure*}[ht]
    \centering
    \includegraphics[width=1.0\textwidth]{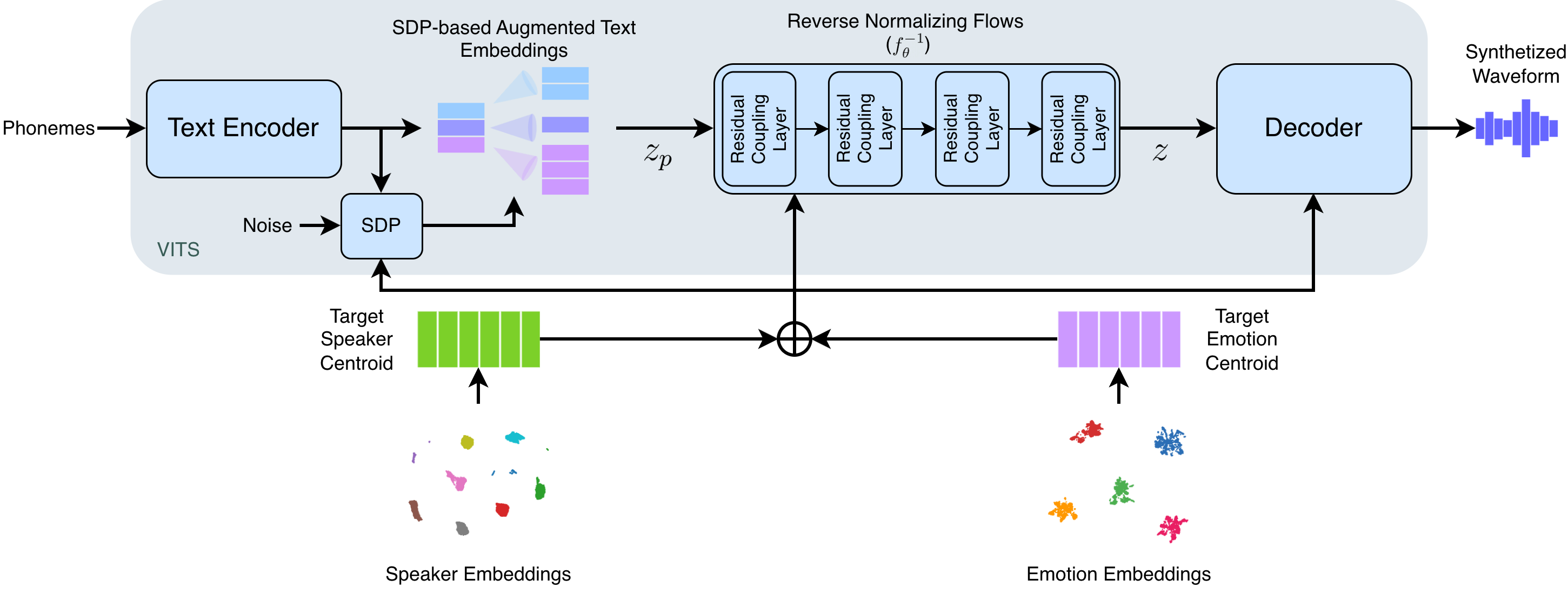}
    \caption{SelfTTS architecture at inference time. The model follows the VITS inference pipeline augmented with disentangled speaker and emotion embeddings. Both embeddings can be obtained either from a reference audio or from post-hoc analysis of the training style space via centroid prototypes.}
    \label{fig:selftts_architecture_infer}
\end{figure*}

As illustrated in Figure~\ref{fig:selftts_architecture_infer}, inference follows the standard VITS pipeline augmented with disentangled speaker and emotion conditioning. The trained SDP upsamples each phoneme embedding to its corresponding frame-level duration, yielding the prior representation $z_p$. The reverse Normalizing Flow then transforms $z_p$ into the latent variable $z$, which is finally decoded into the output waveform by the Decoder.

Both speaker ($g$) and emotion ($e$) embeddings can be obtained either from a reference audio signal at inference time or derived from centroid prototypes computed via post-hoc analysis of the training style space. The two disentangled embeddings are summed and used to condition the SDP, Normalizing Flows, and  Decoder, enabling flexible control over both speaker identity and emotional expressivity at inference.

\subsection{Experimental Setup}

All experiments were conducted on a single NVIDIA L40S GPU with 48GB of VRAM. The model was trained with an initial learning rate of $2 \times 10^{-4}$, utilizing a learning rate decay scheduled by a factor of $0.999^{1/8}$ per epoch. We employed the AdamW optimizer~\cite{loshchilov2018decoupled} with hyperparameters $\beta_1 = 0.8$, $\beta_2 = 0.99$, and a weight decay of $\lambda = 0.01$. The batch size was set to 64. 

To expedite the initial training phases where the model must learn phonetic alignment and waveform generation, we initialized our training from a checkpoint pre-trained on the VCTK dataset~\cite{veaux2017cstr} for 800k steps. This base checkpoint was trained using a standard lookup table to control speaker voice and only the layers compatible with the proposed architecture were transferred. From this initialization, the proposed model was trained for an additional 200k steps, followed by 50k steps of refinement using Self-Augmentation with a reduced learning rate of $2 \times 10^{-5}$. On average, each experimental run required approximately two days to complete.

In order to generate the test sentences, we extracted centroid prototypes~\cite{kwon2019emotional} of the emotion and speaker embeddings using the style space of the training data. Therefore, at inference time each centroid prototype was used to generate the cross-speaker style transfer test samples.

We evaluated our model against two VITS-based architectures designed for cross-speaker style transfer:

\begin{itemize}
    \item \textbf{E3-VITS~\cite{jung2023e3vits}:} This model proposes batch-permuted style perturbation to enhance style transfer capabilities. It generates synthetic data via a Normalizing Flows block, with the resulting outputs optimized by augmenting the original VITS discriminator losses for those outputs. Since our primary dataset lacks emotional descriptions, that specific component was not used.
    
    \item \textbf{VECL~\cite{gudmalwar2024vecl}:} A state-of-the-art model capable of cross-lingual emotion transfer. It relies on pre-trained emotion and speaker encoders alongside consistency losses for speaker identity, emotion, and intelligibility. These losses ensure that the representations of the generated audio remain consistent with ground-truth audio representations when processed by domain-specific pre-trained models.
\end{itemize}

\subsubsection{Data}

To train the proposed model, we used the ESD dataset~\cite{zhou2021emotional}, which consists of an emotional speech corpus containing 350 parallel utterances spoken by 20 speakers (10 native English and 10 native Mandarin Chinese speakers), covering five emotion categories (neutral, happy, angry, sad, and surprise). In total, the corpus contains more than 29 hours of speech recorded in a controlled acoustic environment using professional equipment, ensuring high recording quality. The dataset was specifically designed for emotional voice conversion research, providing wide lexical variability, speaker diversity, and consistent recording conditions. In the context of this work we only used the English speakers and we selected speakers 11 (male) and 15 (female) as target speakers for the cross-speaker style transfer experiments. For these two speakers, only neutral data was used during model training. Their emotional utterances were used as ground-truth for subsequent evaluations and were not included in any stage of model development or training.

\subsubsection{Evaluation}

To evaluate the performance of the generated audio, we adopted both objective metrics and subjective human assessments to verify perceptual quality. As our main goal is to effectively achieve cross-speaker style transfer is important to emphasize that our evaluation protocol considers evaluating only generated speech for the two neutral speakers (speakers 11 and 15) from ESD.

We evaluated several key pillars of cross-speaker synthesis and style transfer quality:

\begin{itemize}
    \item \textbf{Naturalness:} We utilized UTMOS~\cite{saeki2022utmos} to provide a predicted Mean Opinion Score (MOS) for the synthesized audio;
    \item \textbf{Intelligibility:} We employed the Whisper model~\cite{radford_2023_whisper} to transcribe generated audio, subsequently calculating the Word Error Rate (WER) compared to groundtruth transcriptions to measure intelligibility;
    \item \textbf{Speaker and Emotion Similarity:} To assess the models' ability to maintain target timbre and emotion, we calculated the Speaker Embedding Cosine Similarity (SECS) using Resemblyzer\footnote{Resemblyzer available at: \url{https://github.com/resemble-ai/Resemblyzer}} and the Emotion Embedding Cosine Similarity (EECS)~\cite{oh2025dureflexevc}. For EECS, representations were extracted from the emotion2vec+ large model~\cite{ma2024emotion2vec}\footnote{Emotion2vec+ Large available at: \url{https://huggingface.co/emotion2vec/emotion2vec\_plus\_large}}.
    \item \textbf{Disentanglement and Alignment:} We calculated the Centered Kernel Alignment (CKA)~\cite{kornblith2019similarity} between the training representations of the emotion and speaker encoders to measure the degree of entanglement between the two embeddings. Additionally, we evaluated how well each representation aligned with its respective ground-truth labels by comparing them against its Label Kernel~\cite{cortes2012algorithms}. We refer to this metric as LK-CKA.
\end{itemize}

While the UTMOS, WER, SECS, and EECS metrics evaluate the quality of the generated cross-speaker style transfer samples, the CKA and LK-CKA metrics are used to evaluate speaker and emotion embeddings. CKA is a similarity measure that quantifies how aligned two sets of representations are by comparing their pairwise similarity (kernel) matrices in a way that is invariant to orthogonal transformations and isotropic scaling (low values mean less alignment between representations). LK-CKA aims to compare the representations generated by an encoder to its label kernel, measuring how aligned is the encoder representation to the labels it was designed to be aligned to (higher values mean good alignment). 

For the subjective assessment, we randomly selected samples from each evaluated model and conducted a perception test with 30 native English-speaking participants. Each test followed a Mean Opinion Score (MOS) protocol using a 5-point Likert scale (1: Poor, 5: Excellent). 

The evaluation comprised three distinct tests: naturalness (nMOS), emotion similarity (eMOS), and speaker similarity (sMOS). For each test, we selected 2 audio samples per speaker for each of the 4 emotions ($2 \times 2 \times 4 = 16$ samples) across the four primary models and the ground-truth. In total, each participant evaluated 240 samples ($3 \times 16 \times 5 = 240$).

\section{Results}

Table~\ref{tab:evaluation_results} presents the subjective and objective metric results comparing the ground-truth, proposed model, its version without Self-Augmentation, and the two evaluated baselines.

\begin{table*}[htbp]
\centering
\small
\setlength{\tabcolsep}{4pt}
\caption{Subjective and objective performance comparison between SelfTTS and baseline models. Subjective metrics include a 95\% confidence interval. For this and the subsequent tables, the following visualization pattern will be adopted: $\uparrow$ higher is better. $\downarrow$ lower is better. Best values are \textbf{bolded}, second-best values are \underline{underlined} and proposed model is highlighted in green.}
\label{tab:evaluation_results}
\begin{tabular}{c c c c c c | c c c c}
\toprule
\multirow{2}{*}{\textbf{Model}} &  
\multirow{2}{*}{\textbf{\shortstack{Speaker \\ Encoder}}} &  
\multirow{2}{*}{\textbf{\shortstack{Emotion \\ Encoder}}} &  
\multicolumn{3}{c}{\textbf{Subjective Evaluation}} &  
\multicolumn{4}{c}{\textbf{Objective Evaluation}} \\
\cmidrule(lr){4-6} \cmidrule(lr){7-10}
& & & nMOS$\uparrow$ & eMOS$\uparrow$ & sMOS$\uparrow$ & UTMOS$\uparrow$ & WER$\downarrow$ & SECS$\uparrow$ & EECS$\uparrow$ \\
\midrule

GT & - & - & $3.638 \pm 0.106$ & 3.541 $\pm$ 0.122 & 4.009 $\pm$ 0.107 & 3.6560 & 0.0718 & 0.8297 & 0.9659 \\
\midrule \midrule

E3-VITS & Lookup & StyleSpeech & \textbf{2.763} $\pm$ 0.111 & 2.237 $\pm$ 0.108 & 2.815 $\pm$ 0.116 & \textbf{3.9614} & \underline{0.1222} & 0.8069 & 0.5367 \\
\midrule
VECL & HSAP & CNN-based & 2.707 $\pm$ 0.107 & 2.556 $\pm$ 0.108 & 2.804 $\pm$ 0.114 & \underline{3.7959} & \textbf{0.1146} & 0.7806 & 0.6929 \\
\midrule
\multirow{2}{*}{\shortstack{SelfTTS w/o \\ Self-Aug.}} & \multirow{2}{*}{RE} & \multirow{2}{*}{RE} & \multirow{2}{*}{2.228 $\pm$ 0.105} & \multirow{2}{*}{\underline{2.849 $\pm$ 0.108}} & \multirow{2}{*}{\textbf{3.121} $\pm$ 0.112} & \multirow{2}{*}{3.4899} & \multirow{2}{*}{0.1952} & \multirow{2}{*}{\underline{0.8103}} & \multirow{2}{*}{\textbf{0.8793}} \\
& & & & & & & & & \\
\midrule
\rowcolor{medgreen}
SelfTTS & RE & RE & \underline{2.746 $\pm$ 0.109} & \textbf{2.853 $\pm$ 0.107} & \underline{3.112 $\pm$ 0.114} & 3.6104 & 0.1636 & \textbf{0.8163} & \underline{0.8423} \\

\bottomrule
\end{tabular}
\end{table*}

As observed, SelfTTS achieves the best eMOS score and second-best results for nMOS and sMOS, demonstrating its ability to effectively combine naturalness with emotional and speaker consistency. The version without Self-Augmentation yields strong eMOS and sMOS results but reports the lowest naturalness, highlighting the effectiveness of Self-Augmentation in maintaining cross-speaker style transfer capabilities while significantly improving the naturalness of the generated audio.

In terms of objective metrics, the proposed model achieves the closest results to ground-truth in both SECS and EECS, reflecting stronger speaker and emotion adherence. The higher WER observed for SelfTTS suggests a domain mismatch inherent to the cross-speaker style transfer task: by successfully transferring expressive prosody to neutral speakers, the model introduces acoustic patterns that may deviate from the transcription-friendly characteristics of neutral speech. The baseline models, which exhibit lower emotional adherence, are less affected by this phenomenon, as their outputs remain closer to neutral speech acoustics.

Finally, the higher UTMOS scores observed for the baselines, even surpassing ground-truth values in some cases, are likely a consequence of UTMOS not being specifically designed to evaluate emotional speech. Models that produce less expressive, more neutral-sounding output may be inadvertently rewarded by this metric, which reinforces the importance of using emotion-specific evaluations such as eMOS and EECS alongside general naturalness scores.

The E3-VITS model achieves the highest naturalness in both nMOS and UTMOS. However, it also shows the poorest performance in eMOS (2.237) and EECS (0.5367). An inability to generate the target emotion may lead the model to produce the average emotion of the training set—which in this case is closer to neutral. This results in audio samples that fail to meet the desired emotional target, despite their higher quality and intelligibility.

The VECL model presents intermediate results in both subjective and objective metrics, notably achieving the best WER. This superior intelligibility may be attributed to its consistency losses, one of which specifically targets speech intelligibility. Conversely, the higher WER in the proposed model can be justified by its ability to successfully transfer emotion to neutral speakers, whereas the baselines failed to maintain competitive EECS values. When performing cross-speaker style transfer, a model may generate artifacts because emotional data for those specific speakers was never seen during training. Nevertheless, the intelligibility level remains competitive with the literature~\cite{cho2025diemotts}.

Overall, the proposed model demonstrates robustness in the cross-speaker style transfer task, maintaining competitive naturalness and intelligibility alongside superior stability in emotion and speaker identity. These findings are corroborated by the eMOS, sMOS, SECS, and EECS results for both SelfTTS and its non-augmented variant.

\begin{figure}[htbp]
    \centering
    \includegraphics[width=\columnwidth]{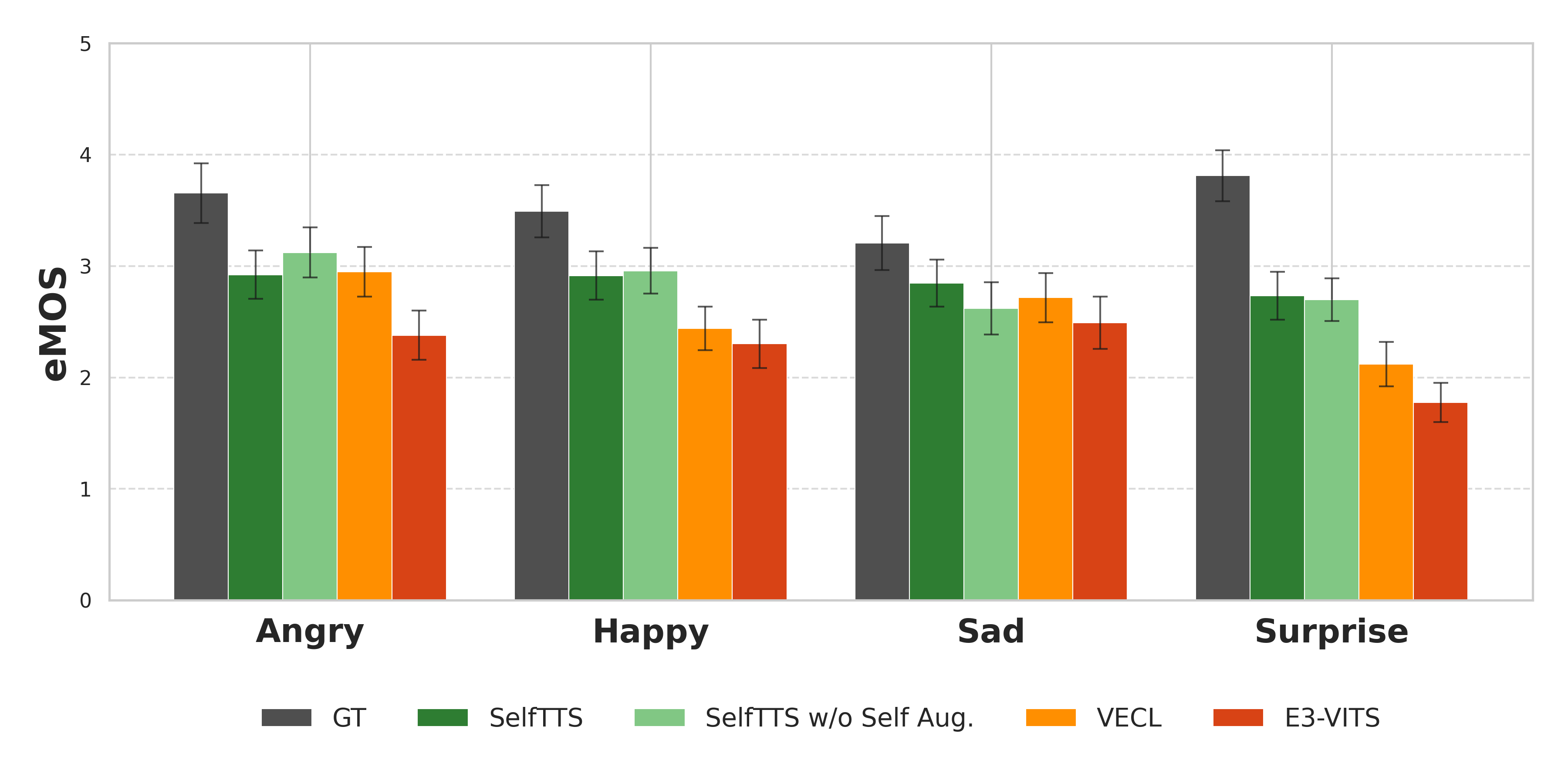}
    \caption{Emotional similarity (eMOS) per emotion category across different models. The models are ordered as: GT (ground-truth), SelfTTS, SelfTTS w/o Self-Aug., VECL, and E3-VITS.}
    \label{fig:emos_per_emotion}
\end{figure}

Figure~\ref{fig:emos_per_emotion} illustrates the eMOS results categorized by emotion for each evaluated model.

We observe that the proposed model maintains higher recognition correspondence across nearly all emotions, particularly in ``Surprise'' and ``Happy'', where it shows the most significant improvement over the baselines. While VECL performed well for the ``Angry'' emotion, its performance was inconsistent across others, whereas SelfTTS exhibits consistency across all emotional categories.

\begin{figure*}[htbp]
    \centering

    % ---------- Top Row ----------
    \subfloat[E3-VITS style space generated by emotion encoder.]{%
        \includegraphics[width=0.38\textwidth]{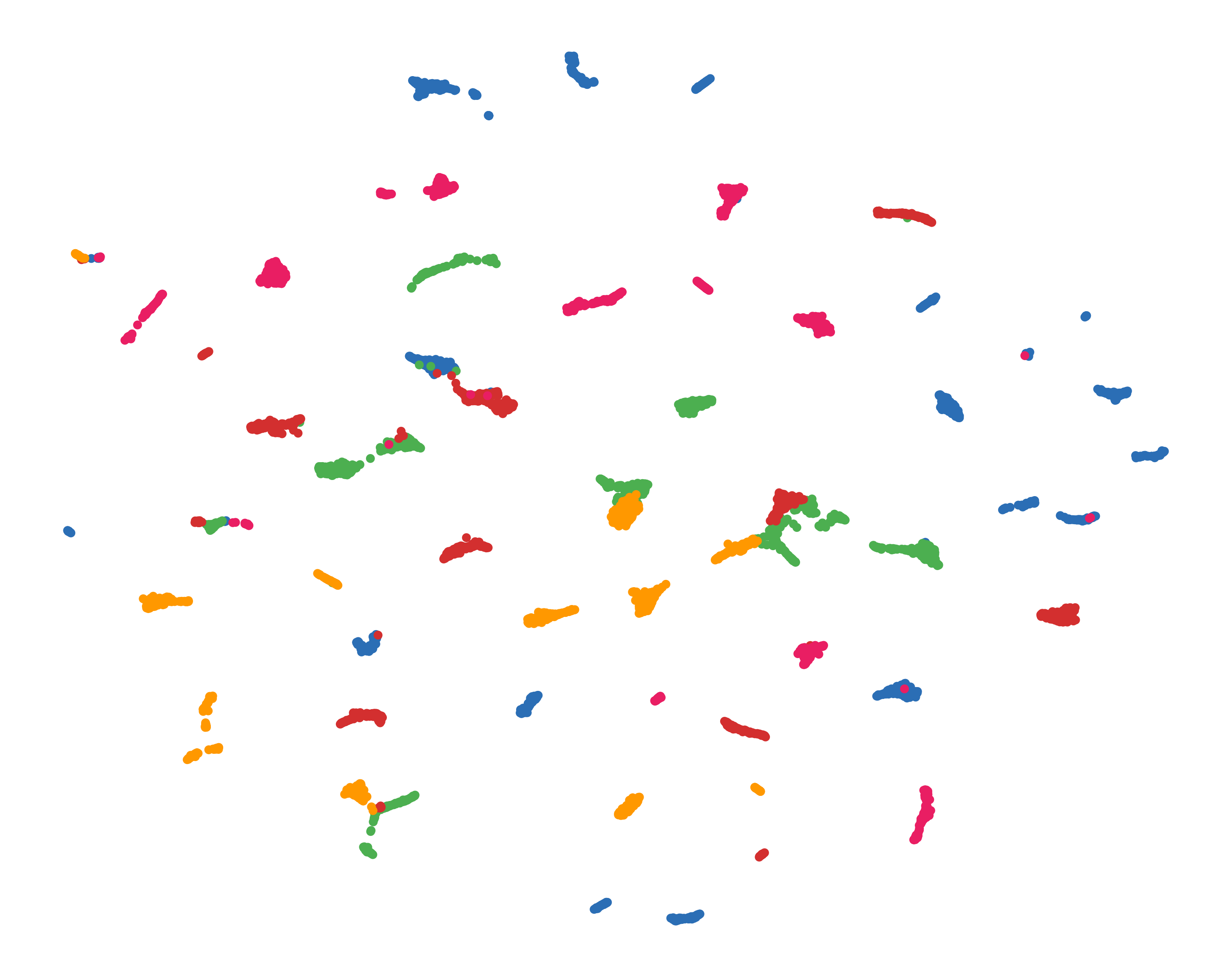}%
    }
    \hspace{0.05\textwidth} 
    \subfloat[VECL style space generated by emotion encoder.]{%
        \includegraphics[width=0.38\textwidth]{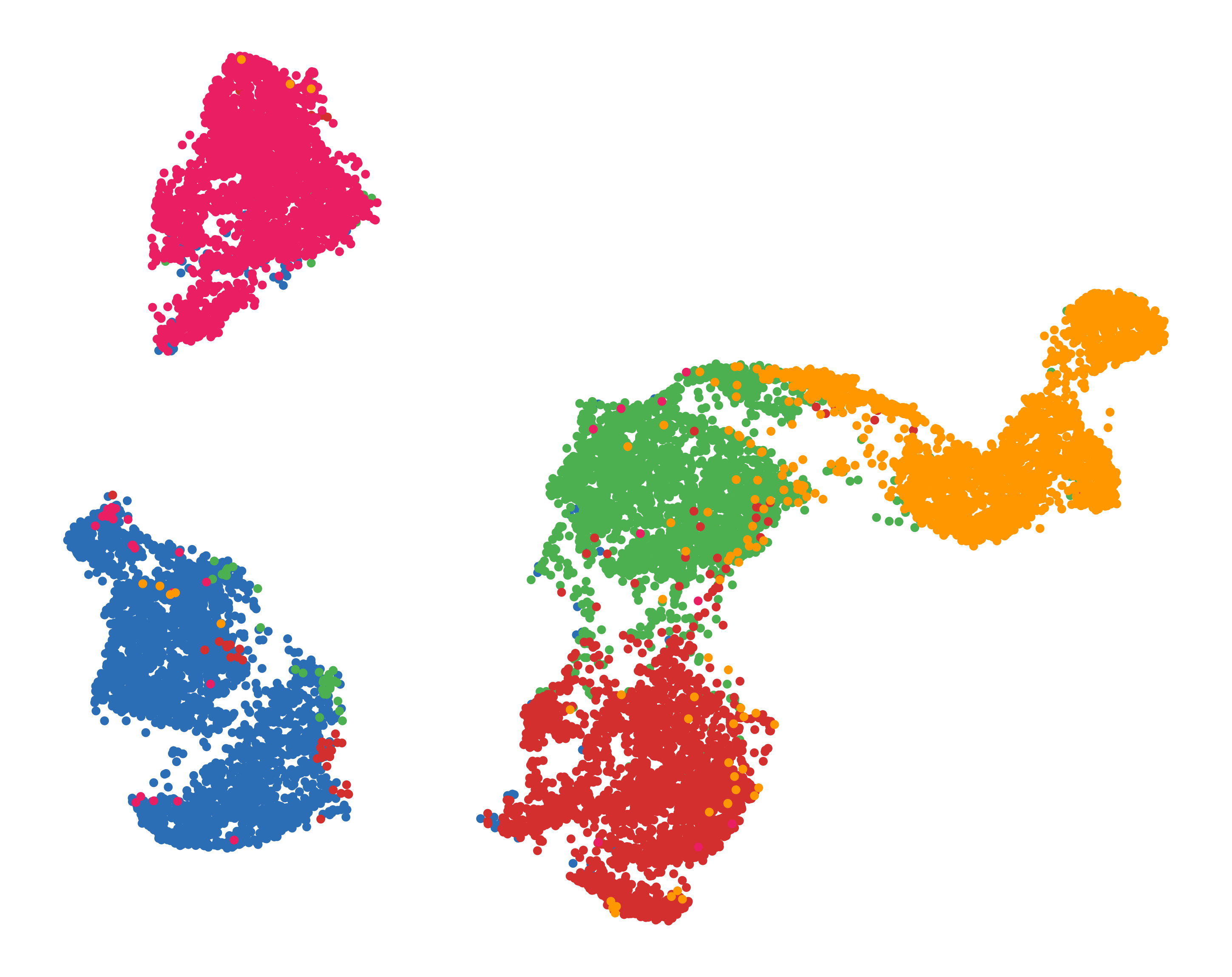}%
    }

    \vspace{0.7cm} % Vertical gap between rows

    % ---------- Bottom Row ----------
    \subfloat[SelfTTS style space generated by emotion encoder.]{%
        \includegraphics[width=0.38\textwidth]{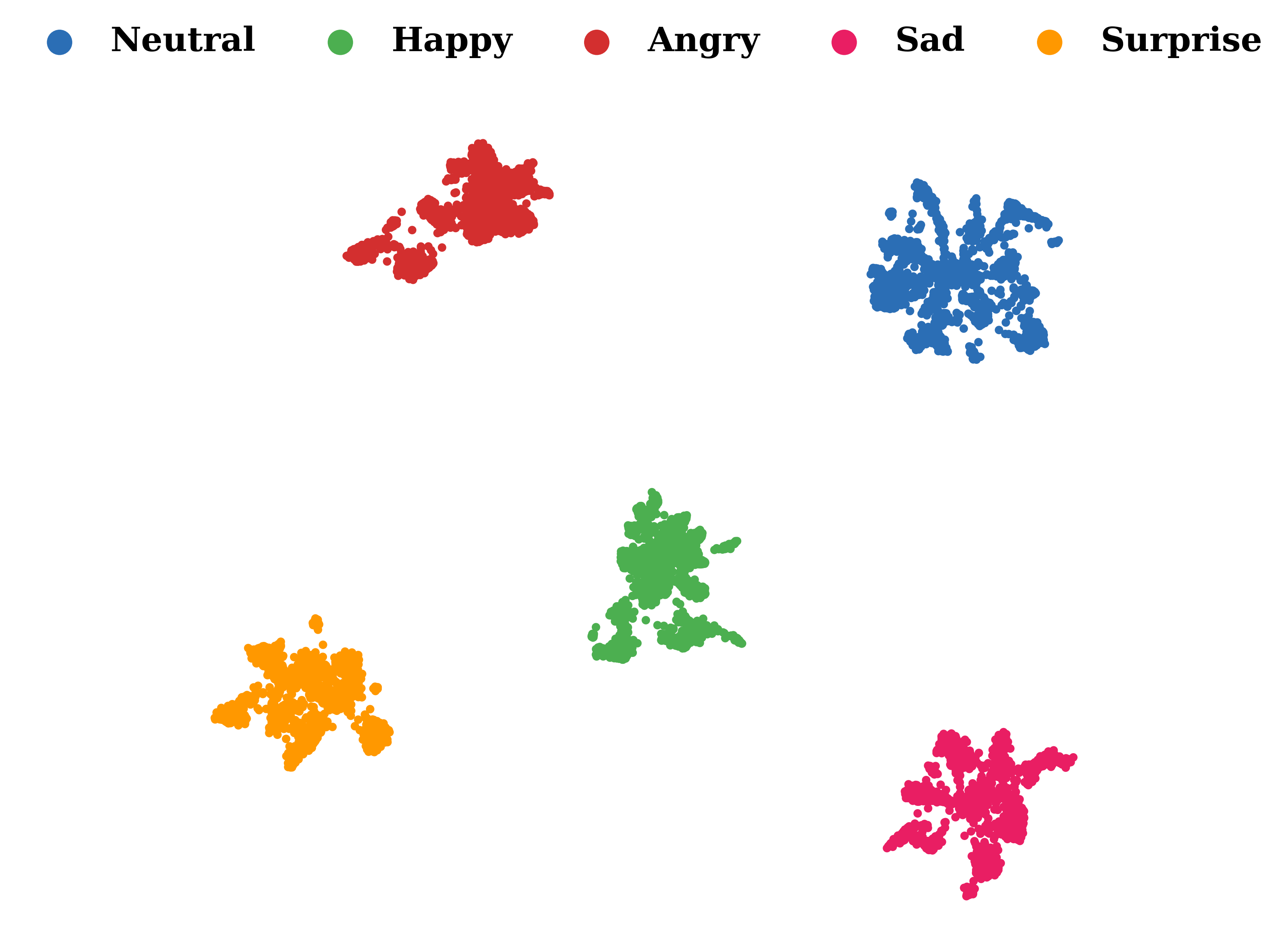}%
    }

    \vspace{0.3cm}
    \caption{UMAP projections of the emotion style spaces for all evaluated models. Embeddings are colored by emotion: Neutral (blue), Happy (green), Angry (red), Sad (pink), and Surprise (orange). The emotion style space of SelfTTS clearly forms well-separated emotional clusters. VECL generates highly concentrated clusters, but there is still overlap between different emotions. E3-VITS is not capable of producing clustered representations.}
    \label{fig:style_space_comparison}
\end{figure*}

To verify the impact of the different emotion encoders used in each model, we extracted the emotion embedding representations and projected them into a 2D style space using UMAP~\cite{mcinnes2018umap} (Figure~\ref{fig:style_space_comparison}).

The proposed model exhibits clear clustering between different emotions which is expected given the MPCL contrastive loss applied. This also justifies its enhanced ability to maintain consistency across different styles during style transfer. VECL shows clear clusters but with visible overlap, while E3-VITS lacks distinct emotional clustering. Without a clear representation between emotions, emotional adherence in generated sample is harder to condition. This aligns with eMOS and EECS performance being lower for E3-VITS and higher for SelfTTS.

\subsection{Towards Developing SelfTTS}

Different architectural configurations and training strategies were explored during the development of SelfTTS. This section presents the experiments conducted, alongside the objective, perceptual, and disentanglement metrics used for selecting the proposed design.

\subsubsection{Loss-based Inductive Bias}

A conventional approach to conditioning emotion and speaker encoder representations to form label-based clusters is the use of Cross-Entropy (CE). However, this loss does not necessarily focus on generating disjoint representations. A more explicit method is contrastive learning. Consequently, we evaluated the impact of using the proposed MPCL contrastive loss versus CE for both emotion and speaker encoders. Furthermore, we assessed the importance of Gradient Reversal Layers (GRL) for disentanglement, comparing its absence against its use with CE and with the proposed cosine-based explicit embedding disentanglement.

\begin{table*}[htbp]
\centering
\small 
\setlength{\tabcolsep}{3pt} 
\caption{Objective metrics of TTS and VC samples generated varying the encoder loss (MPCL $\times$ CE) and Gradient Reversal Layer configuration (None GRL applied, Cross-Entropy loss and proposed Cosine loss). The table also reports speaker-emotion embedding representations similarity (CKA) and speaker and emotion label alignment (LK-CKA).}
\label{tab:ttsvc_results}
\begin{tabular}{c c c c c c c c c c c c c}
\toprule
\multirow{2}{*}{\textbf{\shortstack{Encoder \\ Loss}}} & \multirow{2}{*}{\textbf{GRL}} & \multicolumn{4}{c}{\textbf{TTS}} & \multicolumn{4}{c}{\textbf{VC}} & \multirow{2}{*}{\textbf{CKA}$\downarrow$} & \multicolumn{2}{c}{\textbf{LK-CKA}} \\
\cmidrule(lr){3-6} \cmidrule(lr){7-10} \cmidrule(lr){12-13}
& & UTMOS$\uparrow$ & WER$\downarrow$ & SECS$\uparrow$ & EECS$\uparrow$ & UTMOS$\uparrow$ & WER$\downarrow$ & SECS$\uparrow$ & EECS$\uparrow$ & & \textbf{Speaker}$\uparrow$ & \textbf{Emotion}$\uparrow$ \\
\midrule
CE   & \multirow{2}{*}{None}  & \textbf{3.8585} & \textbf{0.1521} & 0.8064 & 0.7391 & \textbf{3.8324} & \textbf{0.0798} & 0.8120 & 0.7061 & 0.3148 & 0.8704 & 0.6964 \\
MPCL & & 3.6268 & 0.1734 & \textbf{0.8165} & 0.7903 & 3.6968 & \underline{0.0866} & \textbf{0.8213} & 0.7425 & \underline{0.0176} & \textbf{0.9588} & \textbf{0.9648} \\
\midrule
CE   &  \multirow{2}{*}{CE}   & \underline{3.6931} & 0.2046 & 0.8019 & 0.7276 & \underline{3.7239} & 0.0896 & 0.8087 & 0.7043 & 0.0235 & 0.9373 & 0.9282 \\
MPCL &   & 3.4996 & \underline{0.1665} & 0.7704 & 0.7899 & 3.6247 & 0.1013 & 0.7722 & 0.7545 & 0.0336 & 0.9514 & 0.9458 \\
\midrule
CE   & \multirow{2}{*}{Cosine}  & 3.4737 & 0.1775 & 0.8075 & \underline{0.8474} & 3.5507 & 0.0965 & 0.8095 & \underline{0.8107} & 0.0205 & 0.9376 & 0.9024 \\
MPCL &  & \cellcolor{medgreen}3.4899 & \cellcolor{medgreen}0.1952 & \cellcolor{medgreen}\underline{0.8103} & \cellcolor{medgreen}\textbf{0.8793} & \cellcolor{medgreen}3.5503 & \cellcolor{medgreen}0.0899 & \cellcolor{medgreen}\underline{0.8130} & \cellcolor{medgreen}\textbf{0.8530} & \cellcolor{medgreen}\textbf{0.0139} & \cellcolor{medgreen}\underline{0.9581} & \cellcolor{medgreen}\underline{0.9480} \\
\bottomrule
\end{tabular}
\end{table*}

Table~\ref{tab:ttsvc_results} shows that using cosine-based GRL leads to the best EECS results while maintaining strong SECS performance, indicating its efficacy in conveying the correct emotion and keeping the target voice. The CKA calculated between embeddings is the lowest (0.0139), further validating this method for disentanglement, and the LK-CKA for both speaker and emotion are the second best. It is also evident from the results without GRL that using MPCL significantly improves SECS, EECS, and CKA, making it a strong candidate even without explicit embedding disentanglement.

Other metrics remain largely similar, with a negative highlight for CE without GRL, which shows the highest CKA (0.3148) indicating high entanglement and an emotion LK-CKA of 0.6964, suggesting poor alignment with labels. This justifies the lack of visual clustering observed in Figure~\ref{fig:style_space_comparison}.

Notably, audio generated via Voice Conversion (VC) yields results similar to TTS but with better intelligibility (overall average WER of 0.0905 for VC compared to 0.1732 for TTS). This outcome supports the potential of using data generated by this branch for data augmentation, as these samples generally possess better naturalness and intelligibility while maintaining SECS and EECS levels.

\subsubsection{Does Emotion Encoder Input and Architecture Matter?}

Other techniques to avoid speaker and emotion entanglement exist in the literature. One such method involves modifying the emotion encoder input, for example by using only the first 20 mel bins of the spectrogram~\cite{ren2022prosospeech,jiang2023megatts} (where most prosody is concentrated) or using formant shifting which results in a timbre perturbation (TP)~\cite{choi2021neural,lei2022cross,zhu2024metts}. Table~\ref{tab:emotion_encoder_results} presents the results of these techniques and evaluates recent emotion encoders from StyleSpeech~\cite{min2021meta} and StyleTTS~\cite{li2025styletts}. In all cases, the Reference Encoder (RE) was used as the speaker encoder.

\begin{table}[htbp]
\centering
\caption{Objective performance comparison of different emotion encoder configurations and input perturbation methods. $\checkmark$ indicates the feature is used, while $\times$ indicates it is not.}
\label{tab:emotion_encoder_results}
\resizebox{\columnwidth}{!}{%
\begin{tabular}{c c c c c c c}
\toprule
\multirow{2}{*}{\textbf{\shortstack{Emotion \\ Encoder}}} &
\multirow{2}{*}{\textbf{\shortstack{20 Mel \\ Bins}}} &
\multirow{2}{*}{\textbf{TP}} &
\multirow{2}{*}{\textbf{UTMOS}$\uparrow$} &
\multirow{2}{*}{\textbf{WER}$\downarrow$} &
\multirow{2}{*}{\textbf{SECS}$\uparrow$} &
\multirow{2}{*}{\textbf{EECS}$\uparrow$} \\
& & & & & & \\
\midrule
\multirow{4}{*}{RE}
    & \cellcolor{medgreen}$\times$ & \cellcolor{medgreen}$\times$ & \cellcolor{medgreen}3.4899 & \cellcolor{medgreen}0.1952 & \cellcolor{medgreen}0.8103 & \cellcolor{medgreen}\textbf{0.8793} \\
    & $\times$      & $\checkmark$ & 3.5171 & 0.1768 & \textbf{0.8196} & \underline{0.7771} \\
    & $\checkmark$ & $\times$      & 3.5367 & 0.1703 & 0.8102 & 0.7500 \\
    & $\checkmark$ & $\checkmark$ & 3.7208 & 0.2612 & 0.8080 & 0.5933 \\
\midrule
\multirow{4}{*}{StyleSpeech}
    & $\times$      & $\times$      & 3.6500 & 0.1993 & 0.8114 & 0.7621 \\
    & $\times$      & $\checkmark$ & 3.7438 & 0.1676 & 0.8063 & 0.6439 \\
    & $\checkmark$ & $\times$      & 3.7064 & 0.1607 & 0.8160 & 0.6438 \\
    & $\checkmark$ & $\checkmark$ & 3.8258 & \underline{0.1579} & 0.8094 & 0.4816 \\
\midrule
\multirow{4}{*}{StyleTTS}
    & $\times$      & $\times$      & 3.7088 & 0.1591 & \underline{0.8179} & 0.6702 \\
    & $\times$      & $\checkmark$ & \underline{3.8267} & \textbf{0.1516} & 0.8174 & 0.5734 \\
    & $\checkmark$ & $\times$      & 3.8143 & 0.1684 & 0.7985 & 0.5838 \\
    & $\checkmark$ & $\checkmark$ & \textbf{3.8510} & 0.1832 & 0.8000 & 0.5002 \\
\bottomrule
\end{tabular}
}
\end{table}

Notably, the use of both mel filtering and timbre perturbation reduced SECS and EECS, with the best result obtained using the RE without any input perturbation. In particular, emotional adherence (EECS 0.8793) is significantly higher compared to all other experiments. We also note that lower EECS is aligned with higher UTMOS, again suggesting that reduced emotional conditioning leads to higher perceived quality (emotion leakage), which is not desired in this case.

Furthermore, the use of alternative emotion encoders did not positively impact the model, with results falling below the RE while following the same observed pattern as more input perturbation was added. A possible explanation is that explicit embedding disentanglement may benefit from similar architectures to perform more effectively and stabilize the adversarial training, further study would be beneficial.

\subsubsection{Self-Augmentation}

As noted earlier, the voice conversion capability demonstrates strong potential as a data augmentation technique. Tables~\ref{tab:vc_cka_flow_analysis_selftts} and ~\ref{tab:vc_cka_flow_analysis_ce_grl} present the presence of speaker and emotion information at each flow layer during the voice conversion path for the proposed SelfTTS and a baseline utilizing standard Cross-Entropy Gradient Reversal Layer (CE GRL), respectively. The values are calculated via LK-CKA of the mean embeddings at each step. 

\begin{table}[htbp]
\centering
\small
\caption{LK-CKA analysis per module for speaker and emotion label alignment in the VC path (SelfTTS).}
\label{tab:vc_cka_flow_analysis_selftts}
\resizebox{\columnwidth}{!}{%
\begin{tabular}{c c c c}
\toprule
\textbf{Module} & \textbf{Mode} & \textbf{\shortstack{LK-CKA\\Speaker}} & \textbf{\shortstack{LK-CKA\\Emotion}} \\
\midrule
Flow Step 1 & Forward & 0.4830 & 0.0004 \\
Flow Step 2 & Forward & 0.4849 & 0.0006 \\
Flow Step 3 & Forward & 0.2517 & 0.0014 \\
\rowcolor{tableblue} Flow Step 4 & Forward & 0.0270 & 0.0032 \\
Flow Step 5 & Reverse & 0.0916 & 0.1193 \\
Flow Step 6 & Reverse & 0.1273 & 0.1810 \\
Flow Step 7 & Reverse & 0.1309 & 0.2131 \\
Flow Step 8 & Reverse & 0.0918 & 0.2265 \\
Decoder & - & 0.8803 & 0.3307 \\
\bottomrule
\end{tabular}
}
\end{table}

\begin{table}[htbp]
\centering
\small
\caption{LK-CKA analysis per module for speaker and emotion label alignment in the VC path (w/ CE GRL).}
\label{tab:vc_cka_flow_analysis_ce_grl}
\resizebox{\columnwidth}{!}{%
\begin{tabular}{c c c c}
\toprule
\textbf{Module} & \textbf{Mode} & \textbf{\shortstack{LK-CKA\\Speaker}} & \textbf{\shortstack{LK-CKA\\Emotion}} \\
\midrule
Flow Step 1 & Forward & 0.5324 & 0.0005 \\
Flow Step 2 & Forward & 0.5207 & 0.0007 \\
Flow Step 3 & Forward & 0.2692 & 0.0015 \\
\rowcolor{tableblue} Flow Step 4 & Forward & 0.0552 & 0.0030 \\
Flow Step 5 & Reverse & 0.1306 & 0.1218 \\
Flow Step 6 & Reverse & 0.1653 & 0.1703 \\
Flow Step 7 & Reverse & 0.2449 & 0.1843 \\
Flow Step 8 & Reverse & 0.3072 & 0.1798 \\
Decoder & - & 0.9108 & 0.2546 \\
\bottomrule
\end{tabular}
}
\end{table}

Crucially, we observe that the end of the forward flow (Flow Step 4) results in the lowest LK-CKA metrics across both configurations, indicating that the intermediate representation $z_p$ successfully strips away the majority of speaker and emotion information. However, the proposed Explicit Embedding Disentanglement demonstrates a distinct advantage at this bottleneck. While both methods reduce emotion information to near-zero levels at Flow Step 4 (0.0032 for SelfTTS vs. 0.0030 for CE GRL), the proposed method removes approximately 51\% more speaker information compared to the baseline (yielding an LK-CKA of 0.0270 vs. 0.0552).This improved disentanglement effectively yields a "purer" latent $z_p$ that is tightly bound to linguistic content and significantly less entangled with paralinguistic attributes. By more effectively stripping this information during the forward pass, the reverse flow is forced to map $z_p$ to the target latent $z$ relying strictly on the explicit conditioning provided to the network allowing for better cross-speaker style transfer at inference time. 

Interestingly, while speaker information is highly present at the posterior encoder output (Flow Step 1), emotional information is already minimal. During the inverse flow (Flow Steps 5–8), emotion information is gradually reintroduced into the representation, whereas speaker information remains relatively low. This dynamic suggests that other architectural blocks within SelfTTS have assumed the primary responsibility for conditioning speaker identity, allowing the flow modules to focus more exclusively on modeling emotional prosody. Moreover, the Decoder module of SelfTTS has more emotional information that its ablation counterpart, while for speaker information is the opposite. Although this result can be associated with a trade-off between speaker and emotion modeling, results presented in Table~\ref{tab:ttsvc_results} highlights that by using Cosine-based GRL the proposed model achieves better results in terms of both SECS and EECS. 

The promising results in VC path have motivated us to incorporate the synthetic data generated by the model itself during its own training. We evaluated synthetic data in three different configurations: using it as the ground-truth (GT), using it only as new emotional references (ENC) with different speaker's voices, and using it in both (BOTH) configurations. Table~\ref{tab:self_transformation_results} presents the results.

\begin{table}[htbp]
\centering
\caption{Objective performance comparison of different Self-Augmentation configurations.}
\label{tab:self_transformation_results}
\resizebox{\columnwidth}{!}{%
\begin{tabular}{c c c c c}
\toprule
\textbf{Self Aug.} & \multirow{2}{*}{\textbf{UTMOS}$\uparrow$} & \multirow{2}{*}{\textbf{WER}$\downarrow$} & \multirow{2}{*}{\textbf{SECS}$\uparrow$} & \multirow{2}{*}{\textbf{EECS}$\uparrow$} \\
\textbf{config}    &                                          &                          &                          &                          \\
\midrule
GT   & 3.2509  & 0.1784 & 0.7950 & \textbf{0.8896} \\
\rowcolor{medgreen}
ENC  & \textbf{3.6461}  & \underline{0.1554} & \textbf{0.8162} & 0.8027 \\
BOTH & \underline{3.4108} & \textbf{0.1524} & \underline{0.8024} & \underline{0.8425} \\
\bottomrule
\end{tabular}
}
\end{table}

Despite having the lowest EECS among the evaluated configurations, the proposed ENC configuration provides the best speaker adherence (SECS 0.8162) and highest UTMOS (3.6461). Both GT and BOTH configurations showed lower naturalness than the model without Self-Augmentation. Particularly, GT have the lowest impact on improving WER while the other two have improved by around the same level. This lead us to select ENC as the proposed configuration as it presented the more balanced trade-off between improving UTMOS and WER while keeping competitive results in SECS and EECS. The naturalness degradation of using synthetic samples as ground-truth (GT and BOTH configurations) may be due to artificial artifacts that VC is adding to generated samples, forcing the SelfTTS to fit not only the acoustic patterns of ground-truth targets waveforms but also the synthetic artifacts generated by VC.

Finally, we evaluated the impact of varying the masked batch proportion with Self-Augmentation (Table~\ref{tab:aug_proportion_results}).

\begin{table}[htbp]
\centering
\caption{Impact of different Self-Augmentation proportions on objective metrics.}
\label{tab:aug_proportion_results}
\resizebox{\columnwidth}{!}{%
\begin{tabular}{c c c c c}
\toprule
\textbf{Self Aug.} & \multirow{2}{*}{\textbf{UTMOS}$\uparrow$} & \multirow{2}{*}{\textbf{WER}$\downarrow$} & \multirow{2}{*}{\textbf{SECS}$\uparrow$} & \multirow{2}{*}{\textbf{EECS}$\uparrow$} \\
\textbf{proportion} & & & & \\
\midrule
0.00 & 3.5335 & \textbf{0.1538} & 0.8150 & \textbf{0.8778} \\
\rowcolor{medgreen}
0.25 & 3.6104 & 0.1636 & \underline{0.8163} & \underline{0.8423} \\
0.50 & 3.6461 & \underline{0.1554} & 0.8162 & 0.8027 \\
0.75 & \underline{3.6738} & \textbf{0.1538} & \textbf{0.8226} & 0.7651 \\
1.00 & \textbf{3.7663} & 0.1564 & 0.8155 & 0.6624 \\
\bottomrule
\end{tabular}
}
\end{table}

A proportion of 0.00 indicates a continuation of original training for 50k steps with a lower learning rate. Results show that simply continuing training improves naturalness while maintaining high emotional transfer capability. A proportion of 0.25 offered the best balance between naturalness and high SECS/EECS adherence, therefore being selected as the proposed approach. We further observed that increasing synthetic data samples enhances naturalness but negatively impacts EECS, reinforcing the inverse relationship between emotional conditioning and naturalness quality.

\subsection{Cross-corpus}

Transferring style from cross-corpus settings is a harder problem as other acoustic cues may be present in different recording conditions which are not directly related to speaker, emotion or content. This makes the cross-speaker style transfer task even harder. To evaluate cross-corpus capability, we trained SelfTTS with approximately 350 neutral sentences from LJSpeech and VCTK speakers p226 and p231. We performed style transfer inference for these speakers and evaluated them using the objective metrics (Table~\ref{tab:self_tts_speaker_results}). For SECS and EECS, we used the centroid of the respective emotion and speaker embeddings as ground-truth.

\begin{table}[htbp]
\centering
\caption{Objective performance comparison across LJspeech, p226 and p231 speakers from LJspeech and VCTK in cross-corpus scenario for SelfTTS.}
\label{tab:self_tts_speaker_results}
\resizebox{\columnwidth}{!}{%
\begin{tabular}{c c c c c c}
\toprule
\textbf{Model} & \textbf{Speaker} & \textbf{UTMOS}$\uparrow$ & \textbf{WER}$\downarrow$ & \textbf{SECS}$\uparrow$ & \textbf{EECS}$\uparrow$ \\
\midrule
\multirow{3}{*}{\shortstack{SelfTTS w/o \\ Self Aug.}} & p226 & 3.1898 & 0.5464 & \underline{0.8964} & \underline{0.6583} \\
                         & p231 & 3.1937 & 0.5358 & 0.8714 & \textbf{0.6890} \\
                         & LJ   & 3.1800 & \underline{0.3482} & 0.8355 & 0.6133 \\
\midrule
\multirow{3}{*}{SelfTTS} & \cellcolor{medgreen}p226 & \cellcolor{medgreen}\underline{3.3806} & \cellcolor{medgreen}0.5176 & \cellcolor{medgreen}\textbf{0.9114} & \cellcolor{medgreen}0.6074 \\
                         & \cellcolor{medgreen}p231 & \cellcolor{medgreen}\underline{3.3806} & \cellcolor{medgreen}0.5399 & \cellcolor{medgreen}0.8914 & \cellcolor{medgreen}0.6125 \\
                         & \cellcolor{medgreen}LJ   & \cellcolor{medgreen}\textbf{3.3951} & \cellcolor{medgreen}\textbf{0.3327} & \cellcolor{medgreen}0.8587 & \cellcolor{medgreen}0.6006 \\
\bottomrule
\end{tabular}%
}
\end{table}

While SelfTTS w/o Self Aug. loses some emotional adherence, the version with Self-Augmentation shows significant improvement in naturalness, validating the approach even in cross-corpus scenarios. However, the WER is considerably higher than intra-corpus results which may be caused by both the small amount of samples for those specific speakers and the differences in recording conditions as emotional references are only available in ESD recording conditions. High SECS values indicate that the proposed SelfTTS model is not leaking speaker voices. On the other hand the EECS values are lower compared to previous experiments, which can be justified by the mismatch between the speaker voices used in the ESD-based centroid emotion2vec+ embedding reference and generated samples. The lower values also indicate that in cross-corpus scenarios transferring emotions to neutral speakers is even harder.

\section{Conclusion}

In this work, we presented SelfTTS, a self-contained and highly robust text-to-speech model designed to achieve high-quality cross-speaker style transfer. By integrating a novel explicit disentanglement strategy utilizing Gradient Reversal Layers (GRL) combined with a cosine similarity loss, the architecture successfully decouples speaker identity from emotional prosody. Crucially, SelfTTS achieves this without relying on external, large-scale pre-trained speaker or emotion encoders, making it lightweight and structurally independent. 

Our proposed Multi Positive Contrastive Learning (MPCL) loss effectively enforces label-oriented clusters for both speaker and emotion embeddings without requiring complex batching pipelines. Furthermore, we demonstrated that leveraging the model's inherent voice conversion capabilities for a targeted self-refinement process (Self-Augmentation) improves the naturalness and intelligibility of synthesized speech while maintaining strong emotional adherence and target speaker voice consistency. UMAP projections and subjective evaluations corroborate these findings, with SelfTTS establishing superior perceptual emotional similarity (eMOS) and robust stability in target timbre compared to state-of-the-art baselines.

A fundamental pillar of this contribution is its focus on open science and full reproducibility. SelfTTS relies exclusively on publicly available datasets, and both our complete implementation and training configurations have been made entirely open-source. 

Despite these contributions, notable limitations remain. Training on relatively small-scale academic corpora introduces data sparsity that negatively impacts the model's cross-corpus generalization, resulting in elevated Word Error Rates (WER) when testing on unseen acoustic environments like LJSpeech and VCTK. Furthermore, while Self-Augmentation significantly bridges the quality gap, a persistent trade-off remains between high-intensity emotional expressivity and absolute waveform naturalness, leaving room for further refinement in speech synthesis artifact reduction.

\subsection{Future Work}

Future research will focus on the following areas to further refine SelfTTS:

\begin{itemize}
    \item \textbf{Robustness to cross-corpus scenarios:} as maintaining identical recording conditions is often unfeasible in real-world applications, developing models robust to such acoustic differences is a promising direction;
    \item \textbf{Scalability for zero-shot:} Further investigation is required to determine whether the model scales well with more data even with limited amount of expressive samples to perform zero-shot style transfer for unseen speakers;
    \item \textbf{Cross-lingual:} Given the baseline capabilities of models like VECL in cross-lingual settings and the better intelligibility of VC samples, we plan to evaluate the scalability of the MPCL and Self-Augmentation strategies for multilingual expressive TTS.
\end{itemize}

% \section{Acknowledgments}

% \ifcameraready
%      This study is partially funded by CAPES – Finance Code 001. It is also supported by FAPESP (BI0S \#2020/09838-0 and Horus \#2023/12865-8). Paula D. P. Costa, Lucas H. Ueda, João G.T. Lima and Pedro R. Corrêa are affiliated with the Dept. of Computer Engineering and Automation (DCA), Faculdade de Engenharia Elétrica e de Computação, and are part of the AI Lab., Recod.ai, Institute of Computing, UNICAMP. This project was supported by MCTI under Law 8.248/1991, PPI-Softex, published as Cognitive Architecture (Phase 3), DOU 01245.003479/2024-1.
% \else
%      \textit{Anonymized for blind review}.
% \fi

\section{Generative AI Use Disclosure}

Generative AI tools were used during the preparation of this manuscript exclusively for polishing language for clarity and assisting with LaTeX formatting of results. The authors affirm that the core research, methodology, and the manuscript were produced by the human authors, who remain fully responsible and accountable for the work. All authors have consented to this submission.

\bibliographystyle{IEEEtran}
\bibliography{mybib}

\begin{IEEEbiography}[{\includegraphics[width=1in,height=1.25in,clip,keepaspectratio]{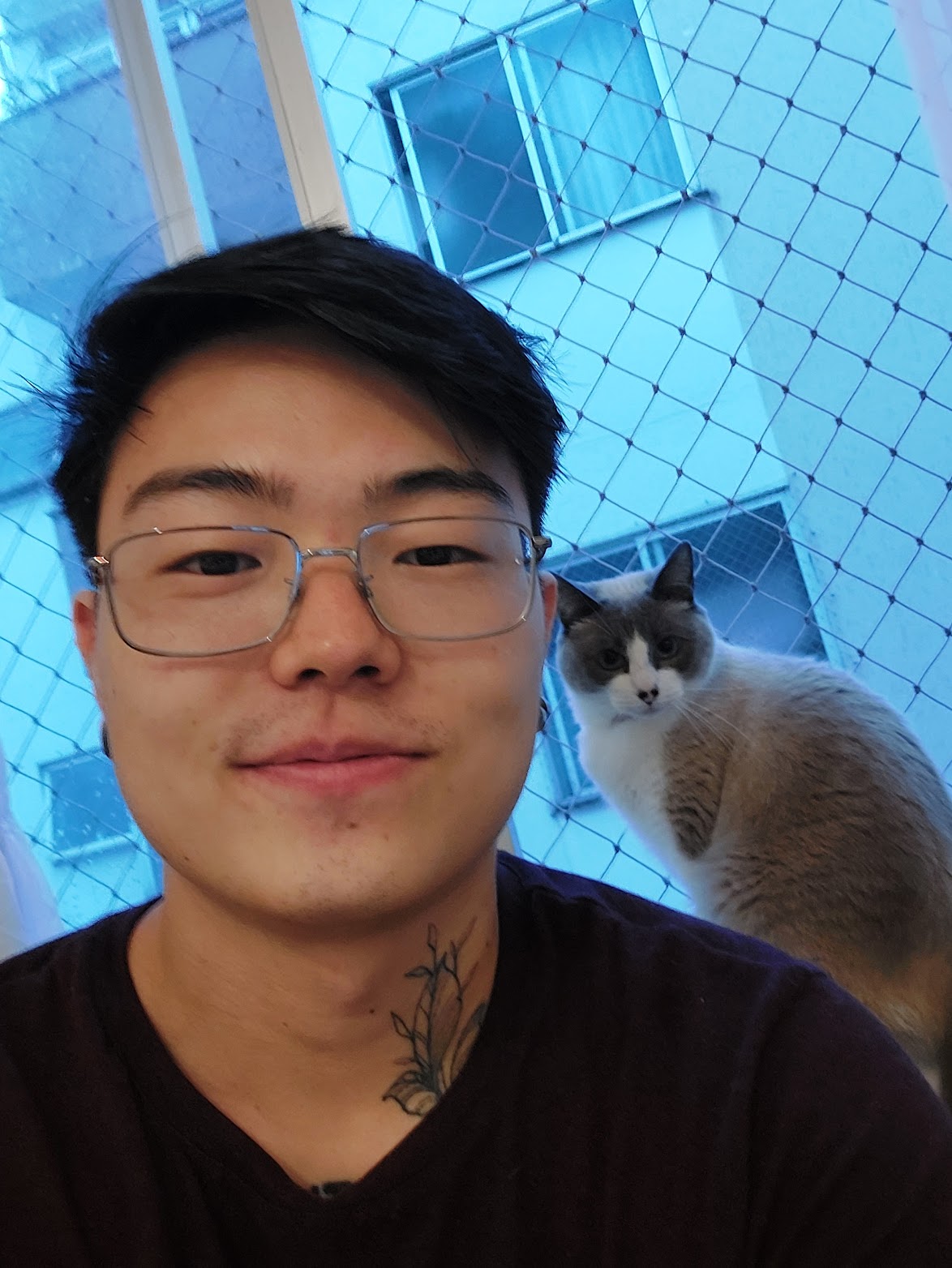}}]{Lucas H. Ueda}
is a Ph.D. Student in Computer Engineering at Universidade Estadual de Campinas (UNICAMP). He earned a
M.Sc. in Computer Engineering from UNICAMP (2021) after a B.Sc. in Applied Mathematics from UNICAMP (2019). His main research relies on Expressive Speech Synthesis with focus on cross-speaker style transfer. Research interests include
expressive text to speech, speech emotion recognition, affective computing, representation learning and disentanglement techniques.
\end{IEEEbiography}

\begin{IEEEbiography}[{\includegraphics[width=1in,height=1.25in,clip,keepaspectratio]{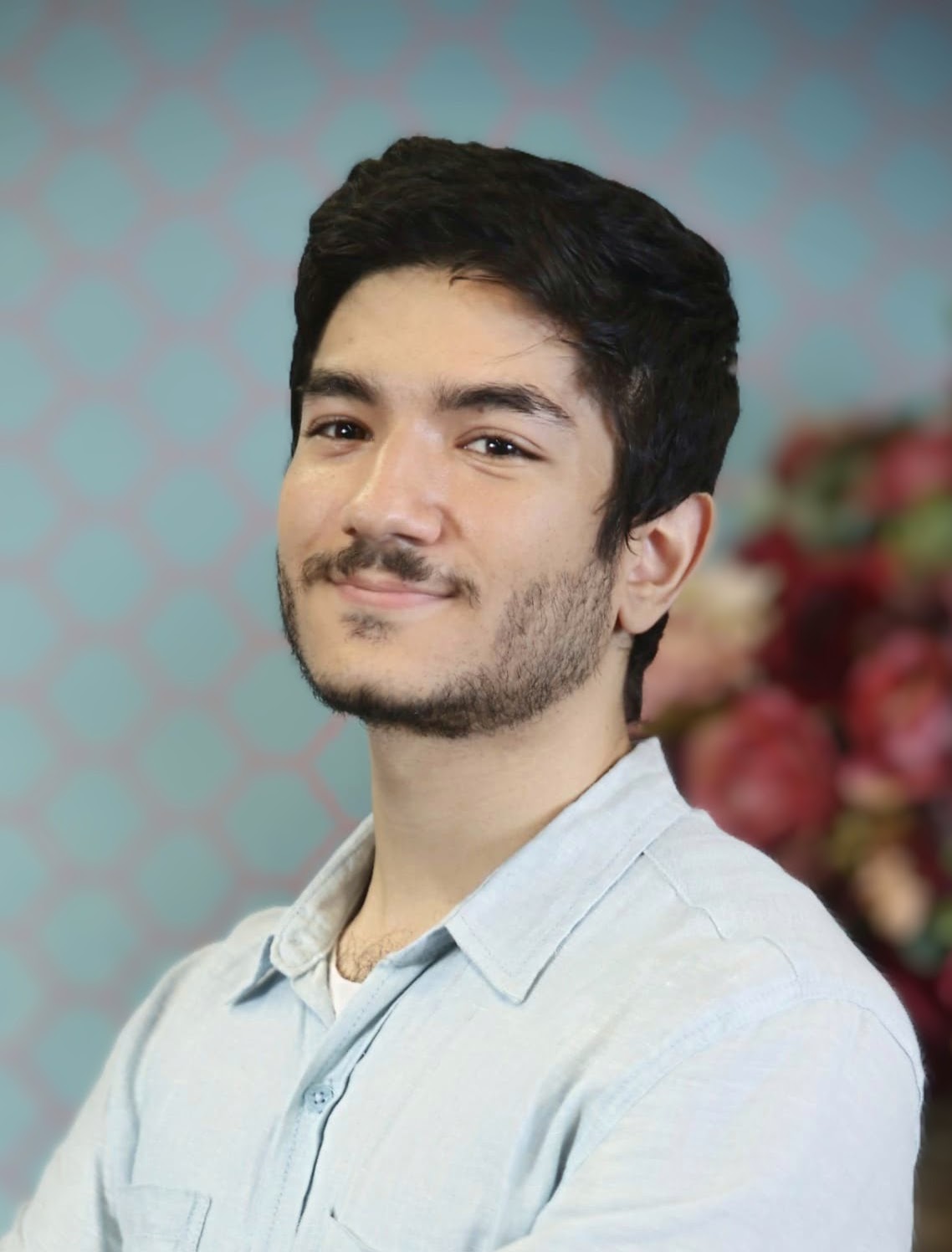}}]{João G.T. Lima} is a M.Sc. student in Computer Engineering at Universidade Estadual de Campinas (UNICAMP), and has completed a research internship at KTH Royal Institute of Technology. He holds a Bachelor’s degree in Linguistics from UNICAMP (2023). His research interests lie at the intersection of phonetic science and speech technologies, with a focus on speech synthesis, conversational systems, and perceptually grounded feature modeling.
\end{IEEEbiography}

\begin{IEEEbiography}[{\includegraphics[width=1in,height=1.25in,clip,keepaspectratio]{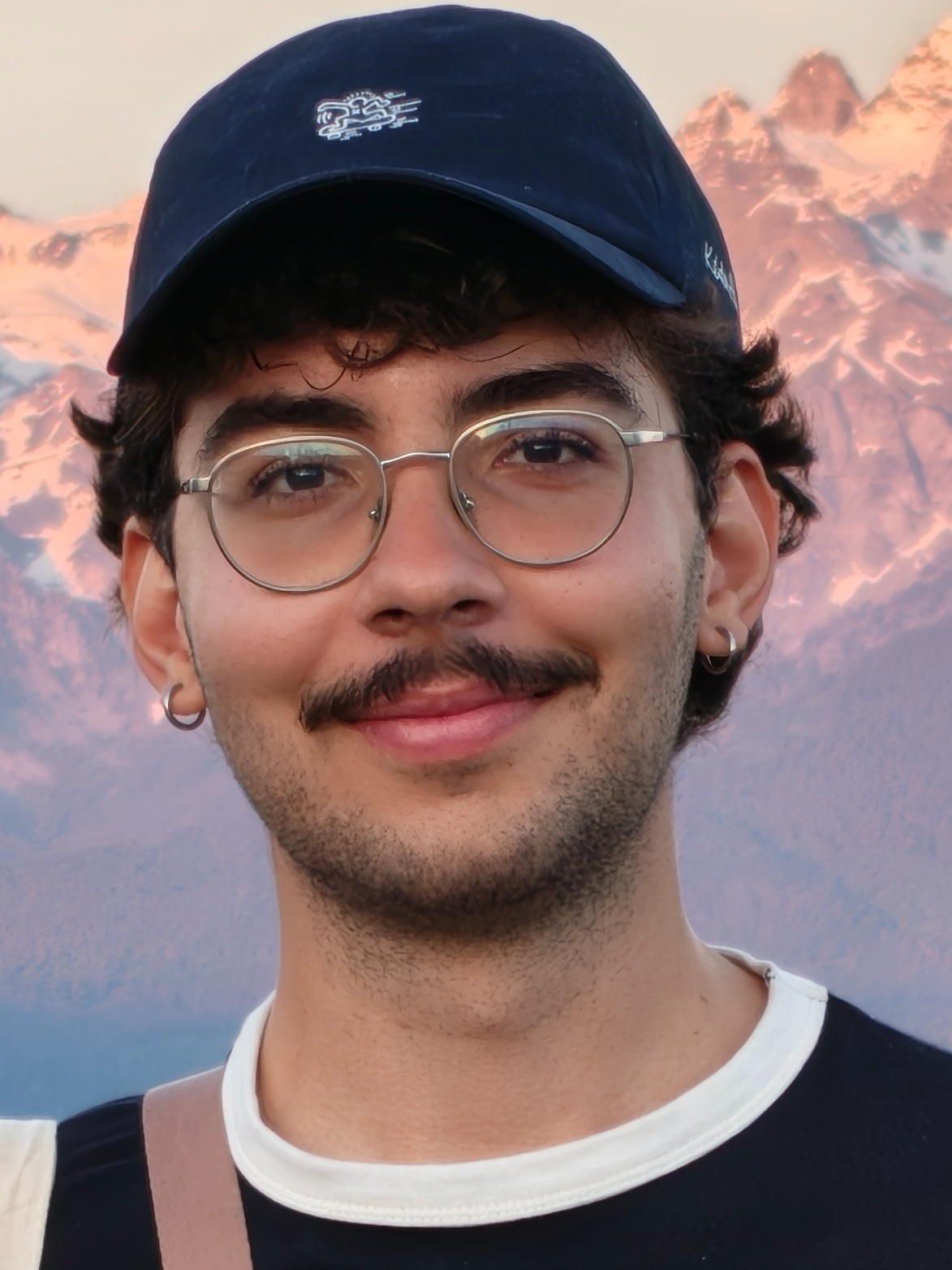}}]{Pedro R. Corrêa} is a M.Sc. student in Computer Engineering at Universidade Estadual de Campinas (UNICAMP), and has earned a B.Sc. in Physics from UNICAMP in 2023. He is about to start his Ph.D. within the collaboration between GIPSA-lab (Université Grenoble Alpes) and Inria. Research interests include affective computing, facial expressions, human speech and conversational systems.

\end{IEEEbiography}

\begin{IEEEbiography}[{\includegraphics[width=1in,height=1.25in,clip,keepaspectratio]{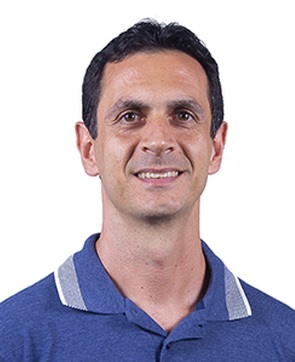}}]{Flávio O. Simões} is a Researcher at CPQD. He earned an M.Sc. in Electrical Engineering from the Universidade Estadual de Campinas - UNICAMP (1999), after a B.Sc. in Computer Engineering from UNICAMP (1996). His research focuses mainly on speech technologies such as speech synthesis, voice biometrics brain-to-speech and conversational systems. Research interest also include machine learning, deep learning and generative AI.
\end{IEEEbiography}

\begin{IEEEbiography}[{\includegraphics[width=1in,height=1.25in,clip,keepaspectratio]{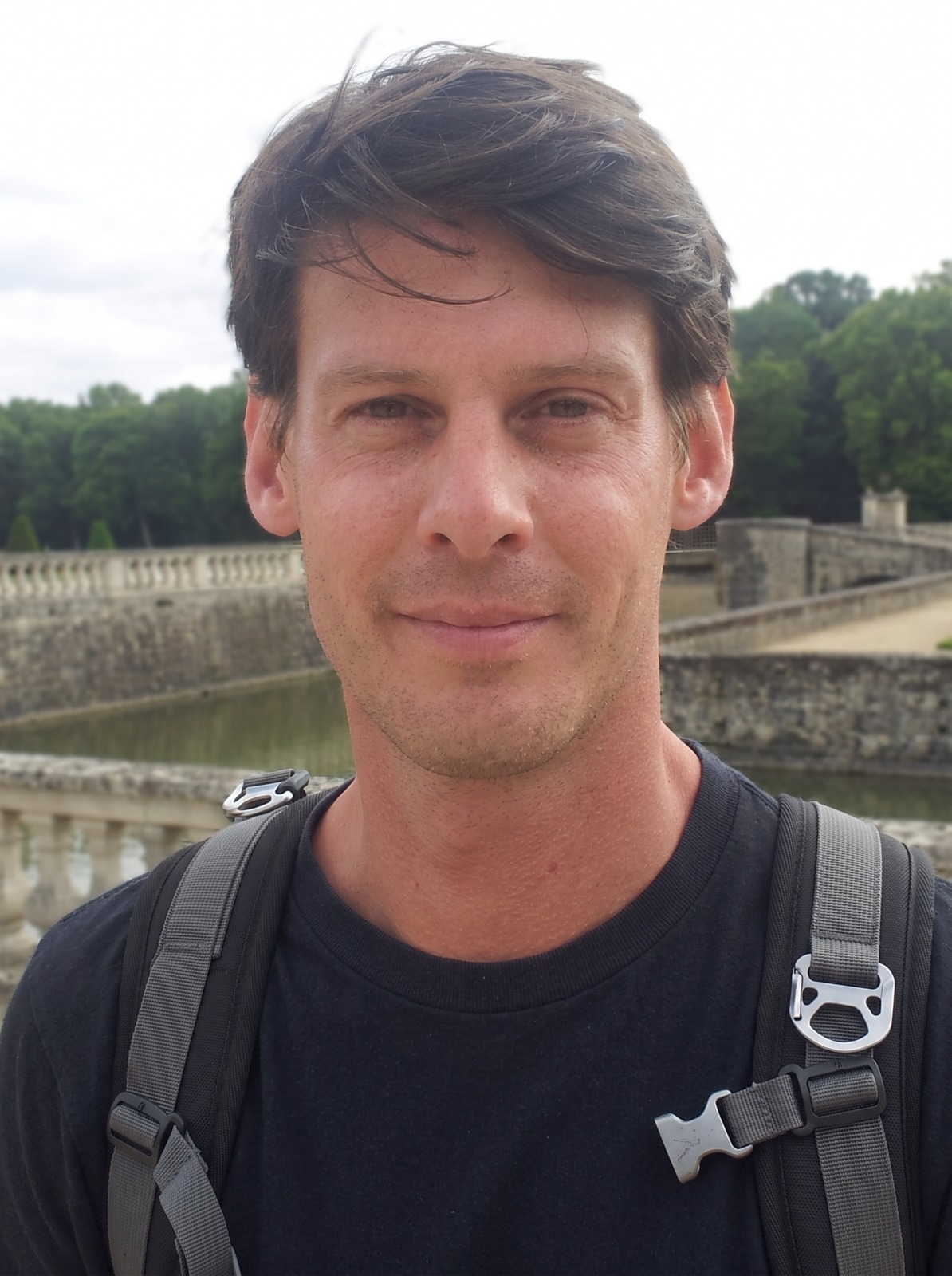}}]{Mário U. Neto} is a researcher at CPQD. He earned an M.Sc. in Electrical Engineering from the Universidade Estadual de Campinas (UNICAMP). His research focus mainly on speech technologies such as speech synthesis, voice conversion, speech recognition, voice biometrics, and speech coding. Research interest also include Artificial Intelligence, machine learning, and Generative AI.
\end{IEEEbiography}

\begin{IEEEbiography}[{\includegraphics[width=1in,height=1.25in,clip,keepaspectratio]{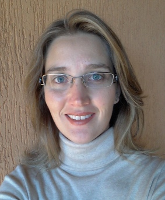}}]{Paula D. P. Costa} is a Professor in the School of Electrical and Computer Engineering (FEEC) at the Universidade Estadual de Campinas (UNICAMP), Brazil. She received her Ph.D. in Computer Engineering from UNICAMP in 2015. Her research focuses on multimodal and generative artificial intelligence, affective computing, and cognitive architectures, particularly for socially interactive agents and robotics.
\end{IEEEbiography}

\vfill

\end{document}